\documentclass[twocolumn]{aastex63}
\usepackage{graphics,epsf}
\usepackage{amsmath}                
\usepackage{amsfonts}               
\usepackage{amssymb}                
\usepackage{epsfig}                 
\usepackage{appendix}
\usepackage{graphicx}
\usepackage{float}
\usepackage{color}
\usepackage{multirow}
\usepackage{colortbl}
\usepackage[para,online,flushleft]{threeparttable}

\newcommand{\cm}{{~\rm cm}}

\newcommand{\km}{{~\rm km}}
\newcommand{\s}{{~\rm s}}

\newcommand{\g}{{~\rm g}}

\newcommand{\K}{{~\rm K}}
\newcommand{\erg}{{~\rm erg}}
\newcommand{\yr}{{~\rm yr}}

\newcommand{\days}{{~\rm days}}

\usepackage{lineno}

\begin{document}

\title{Faint intermediate luminosity optical transients (ILOTs) from engulfing exoplanets on the Hertzsprung gap}


\author{Omer Gurevich}
\affiliation{Department of Physics, Technion, Haifa, 3200003, Israel; ealealbh@gmail.com;; soker@physics.technion.ac.il}

\author{Ealeal Bear}
\affiliation{Department of Physics, Technion, Haifa, 3200003, Israel; ealealbh@gmail.com;; soker@physics.technion.ac.il}

\author{Noam Soker}
\affiliation{Department of Physics, Technion, Haifa, 3200003, Israel; ealealbh@gmail.com;; soker@physics.technion.ac.il}

\begin{abstract}
We follow the evolution of four observed exoplanets to the time when the respective parent star of each planet evolves off the main sequence and engulfs its planet to start a common envelope evolution (CEE), concluding that in each case this process powers an intermediate luminosity optical transient (ILOT; luminous red nova). We characterise the final thousands of days of the orbital decay towards a CEE and determine the properties of the star at the  onset of the CEE. We scale the properties of the ILOT V1309~Scorpii to the properties of a planet that enters a CEE inside a star on and near the Hertzsprung gap to estimate the duration and luminosity of the expected ILOT. 
Based on these we estimate that for a planet of Jupiter mass the ILOT will last for several days and reach a luminosity of several thousand solar luminosity. This type of ILOTs are less luminous than classical novae. Because of the small amount of expected dust and the small amount of energy that an accretion process onto the planet can release, such ILOTs can teach us on the merger at the onset of CEE of stellar companions. Our study adds to the variety of ILOTs that planets can power as they interact with a more massive companion. 
\end{abstract}


\section{Introduction}
\label{sec:intro}

There is a growing  heterogeneous group of transients with peak luminosities in the general range from luminosities of classical novae to typical luminosities of supernovae that last from days to years (e.g., \citealt{Mouldetal1990, Rau2007, Ofek2008, Masonetal2010, Kasliwal2011, Tylendaetal2011, Tylendaetal2013, Kasliwal2013, Tylendaetal2015, Blagorodnovaetal2017, Kaminskietal2018, Pastorelloetal2018, BoianGroh2019, Caietal2019, Jencsonetal2019, PastorelloMasonetal2019, Banerjeeetal2020, Stritzingeretal2020, Blagorodnovaetal2021, Kaminskietal2021CKVul, Pastorelloetal2021}). Some of these transients are powered by thermonuclear reactions while other are powered by gravitational energy. Gravitational energy includes either a mass transfer in a binary system, possibly with the launching of jets by the mass-accreting star (e.g., \citealt{Kashietal2010, SokerKashi2016TwoI, Kashi2018Galax, Soker2020ILOTjets}), and/or mass ejection in the equatorial plane (e.g., \citealt{Pejchaetal2017, HubovaPejcha2019}), or a merger process of two objects. The merger itself can be a process where one star destroys the other star to form an accretion disk, where the merger of the two stars forms a more massive star, or the merger can form a common envelope evolution where one star spirals-in inside the envelope of the larger star (e.g., \citealt{RetterMarom2003,Tylendaetal2011, Ivanovaetal2013a, Nandezetal2014, Kaminskietal2015, Pejchaetal2016a, Pejchaetal2016b, Soker2016GEEI, MacLeodetal2017, MacLeodetal2018, Michaelisetal2018, Segevetal2019, Howittetal2020, MacLeodLoeb2020, Schrderetal2020}). In some cases both CEE and mass transfer can take place, as in the common envelope jets supernova impostor scenario where a neutron star or a black hole spirals-in inside the envelope of a red supergiant star accretes mass and launches jets \citep{SokerGilkis2018, Gilkisetal2019, YalinewichMatzner2019, Schreieretal2021}. 

There is no nomenclature in consensus for this heterogeneous transient group. We refer to all transients that are powered by gravitational energy as Intermediate Luminosity Optical Transients (ILOTs; \citealt{Berger2009, KashiSoker2016, MuthukrishnaetalM2019}), but note that some do not use this term, e.g., \cite{Jencsonetal2019}. \cite{KashiSoker2016}\footnote{See \url{http://physics.technion.ac.il/~ILOT/} for an updated list.} use one set of sub-classes while  \cite{PastorelloMasonetal2019} and \cite{PastorelloFraser2019} use another set. 

In this paper we concentrate on the formation of a CEE where a planet enters the envelope of a low mass star that is just evolving off the main sequence, i.e., mainly during the Hertzsprung gap. 
There are tens of papers on evolved stars, in particular on their giant branches, that swallow planets (e.g., \citealt{Soker1996, SiessLivio1999AGB, Massarotti2008, Carlbergetal2009, VillaverLivio2009, MustillVillaver2012, Nordhausetal2010, NordhausSpiegel2013, GarciaSeguraetal2014, Staffetal2016, AguileraGomezetal2016, Geieretal2016, Guoetal2016, Priviteraetal2016, Veras2016, Raoetal2018, SabachSoker2018a, SabachSoker2018b, Guidarellietal2019, Schaffenrothetal2019, Hegazietal2020, Jimenezetal2020, Krameretal2020, Chamandyetal2021, Guidarellietal2021, Lagosetal2021, Merlovetal2021}). In studying specific observed exoplanetary systems we follow earlier such studies  (e.g., \citealt{NordhausSpiegel2013, SabachSoker2018a, Hegazietal2020}), although these concentrated on the giant branches. We focus on engulfment events at earlier evolutionary phases that lead to ILOTs. 

There are several earlier studies of ILOTs powered by star-planet interaction (for a recent list see \citealt{Kashietal2019Galax}). \cite{RetterMarom2003} and \cite{Retteretal2006} suggested that planets that entered the envelope of a young star powered the ILOT V838~Mon. They suggested this star-planets interaction scenario as an alternative to the stellar merger scenario of V838~Mon \citep{SokerTylenda2003, Tylendaetal2005, Kaminskietal2021V838Mon}. 
\cite{Bearetal2011} discuss ILOTs that result from the destruction of a planet
by a brown dwarf. These ILOTS will be of only several days duration and their luminosity might be even below those of classical novae. We here also deal with such faint ILOTs.  \cite{KashiSoker2017planet} proposed that
the unusual outburst of the young stellar object ASASSN-15qi was a result of a young stellar object that tidally destroyed a sub-Jupiter young
planet. The gas from the destroyed planet formed an accretion disc around the young stellar object, and the accretion process powered this outburst. 
\cite{Kashietal2019Galax} suggested that the $\approx 800$ days long eruption of the young stellar object ASASSN-13db was powered by a young stellar object that accreted the remains of a planet that it shredded by tidal forces. 

We here use the evolutionary code \textsc{mesa}-binary (section \ref{sec:Numeical}) to examine the engulfment of observed exoplanets by their parent stars as the stars just leave the main sequence and have radii of several solar radii (section \ref{sec:Evolution}), so that the outbursts might be large enough to be observed (section \ref{sec:ILOTs}). 
We summarise our results in section \ref{sec:Summary}. 

\section{Numerical scheme}
\label{sec:Numeical}

We follow the evolution of five exoplanetary systems using the evolutionary stellar code Modules for Experiments in Stellar Astrophysics (\textsc{mesa}; \citealt{Paxtonetal2011, Paxtonetal2013, Paxtonetal2015, Paxtonetal2018, Paxtonetal2019}), version 10398 in its binary mode. 
We follow the example of \textsc{mesa}-binary plus point until a CEE is formed, i.e., until $R_1>a(1-e)$, where $R_1$ is the stellar radius, $a$ is the semi-major axis of the orbit and $e$ is the eccentricity. 
We set the parameters of the binary (in inlist~project) \textit{do\_tidal\_circ} and \textit{do\_tidal\_sync} to be true using the default prescriptions for both. Similar to our earlier study in \cite{Rapoportetal2021}, we take the stellar mass-loss rate on the red giant branch from Reimers \citep{Bascheketal1975} with the mass loss parameter $\eta=0.12$.  We set the initial stellar equatorial velocity to be similar to that of the sun, between $ 1-2 \km \s^{-1}$. To chose  a number, we simply set the initial rotation period to be equal to the planet's orbital period (however, this is just an initial condition, and does not result from tidal synchronisation at this early time). We also simulated one case with a zero equatorial rotation velocity and basically we found the same results.

\section{Evolution to CEE}
\label{sec:Evolution}

We examine the evolution of five exoplanets that we found in the Extrasolar Planets Encyclopaedia; (exoplanet.eu; \citealt{Schneideretal2011}). We list these systems in Table \ref{tab:Table1}. We study in details only the four systems where the star has a mass of $M_{\rm i} > 0.85 M_\odot$, as only such similar exoplanet systems might lead to observe ILOTs.
\begin{table*}[]
\centering
\begin{tabular}{|c|c|c|c|c|c|l|c|c|c|c|c|c|c|}
\hline
 System & $M_*$ & $M_{\rm p}$ & $a_{\rm i}$ & $e_{\rm i}$ & $R_{\rm i}$ & Ref. & $t_{\rm f}$ & $e_{\rm f}$ & $R_{\rm f}$ & $L_{\rm f}$ &  $A$ & $B$ & $t_0$ \\ \hline
\textbf{}  & $M_\odot$ & $M_{\rm J}$ & $R_\odot$ &  &  $R_\odot$ & & $10^9\yr$ &  & $R_\odot$ & $L_\odot$ & day & days & day\\ \hline
V 1298 Tau b & 1.1 & 2.2 & 36.98 & 0.29 & 0.98   & D2019 & 8.43 & 0.23 & 6.78  & 19.65 & 4.23  & 13.38 & 12,170 \\ \hline
TOI-216 c & 0.77 & 0.56 & 34.55 & 0 & 0.7        & D21/N21      & 31.76 & 0 & 5.53        & 12.05 &       &       &        \\ \hline
NGTS-11 b & 0.86 & 0.37 & 35.46 & 0.11 & 0.77    & G2020 & 21.15 & 0.11 & 4.1   &  7.23 & 1.736 & 10.82 & 12,340 \\ \hline
Kepler-145 c & 1.28 & 0.25 & 54.32 & 0.11 & 1.25 &C16/V15& 4.71 & 0.11 & 5.42  & 14.3 & 2.144 & 8.767 & 12,350 \\ \hline
Kepler-89 e & 1.28 & 0.11 & 42.88 & 0.02 & 1.25  & W2013 & 4.71 & 0.02 & 5.26  & 13.78  & 1.788 & 19.88 & 14,820 \\ \hline
\end{tabular}
\caption{The five systems that we examine in this paper, from which we study in details four systems. From left to right we list the exoplanet name, the mass of the star, the mass of the planet, the semi-major axis, and the eccentricity. These are our initial conditions for the simulations. In the sixth column we list the observed stellar radius and in the seventh column we give the reference to these observational data. We then list the final time of our simulation, i.e., the onset time of the CEE relative to the zero age main sequence, the final eccentricity of the orbit, the final radius of the star, and the final luminosity of the star. In the last three columns we list the parameters that we fit to equation (\ref{eq:Orbit}). 
References from top to bottom are as follow. 
D2019: \cite{Davidetal2019}; 
D21/N21: \cite{Dawsonetal2021} and \cite{Nesvornyetal2021};
G2020: \cite{Gilletal2020};
W2013: \cite{Weissetal2013};
C16/V15: \cite{Campanteetal2016} and \cite{VanEylenetal2015}.
}
\label{tab:Table1}
\end{table*}

\begin{figure*}
\hskip -1.00 cm
\includegraphics[trim= 0cm 4.7cm 0.0cm 4.9cm,clip=true,width=0.9\textwidth]{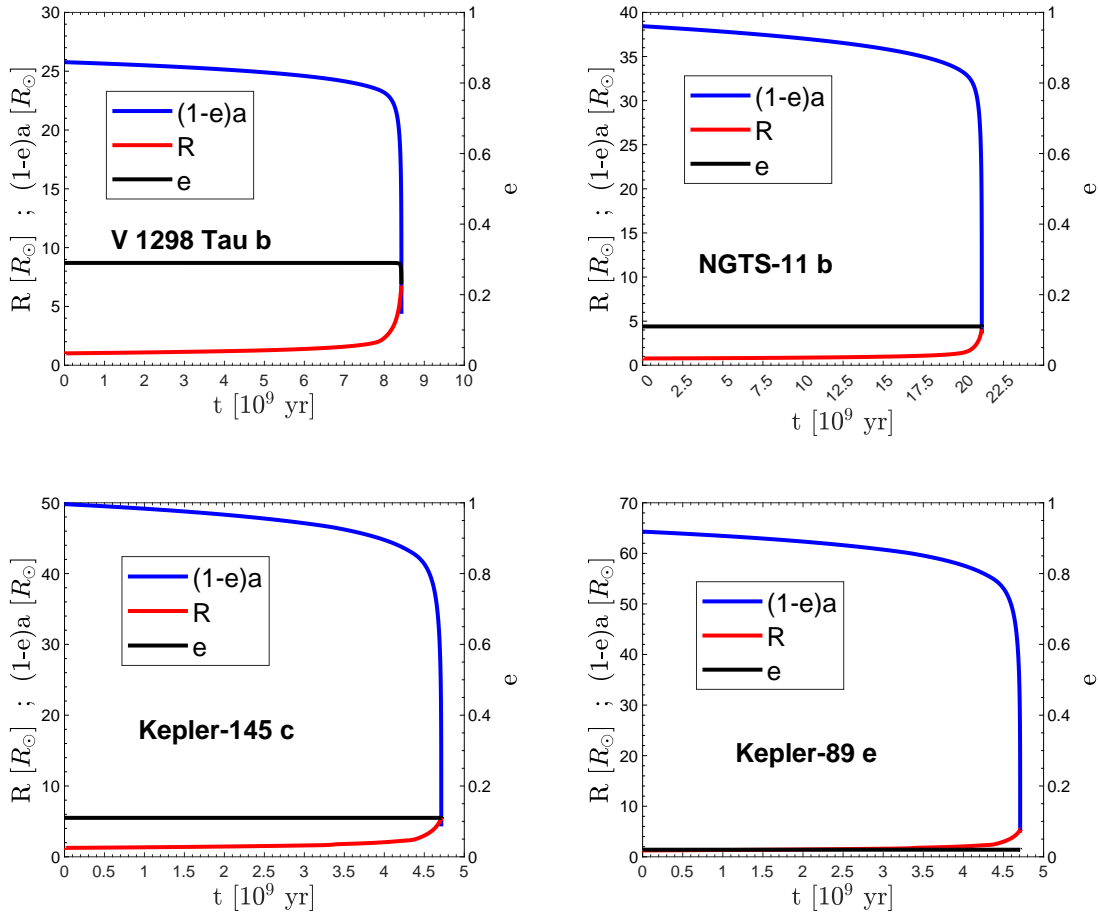}
\vskip +0.00 cm
\caption{The periastron distance $(1-e)a$ (blue line), the eccentricity $e$ (black line scale on the right) and the stellar radius $R$ (red line), as function of time for the four exoplanetary systems that we study here. }
\label{fig:periastron_radius_e_age}
\end{figure*}

The final evolution time of our simulation $t_f$ is at the onset of the CEE, namely, when the periastron distance just becomes smaller than stellar radius, i.e., $a_{\rm f} (1-e_{\rm f}) \le R_{\rm f}$. In Fig. \ref{fig:periastron_radius_e_age} we present the periastron distances and the stellar radii as function of time from the main sequence to the CEE for the four systems we study here.
The evolution is typical for stars that engulf exoplanet (e.g., \citealt{Hegazietal2020}), i.e., a slow change with a final rapid drop towards engulfment and the formation of a CEE. 

\begin{figure}
\includegraphics[trim= 7.0cm 4.5cm -0cm 5.0cm,clip=true,width=0.85\textwidth]{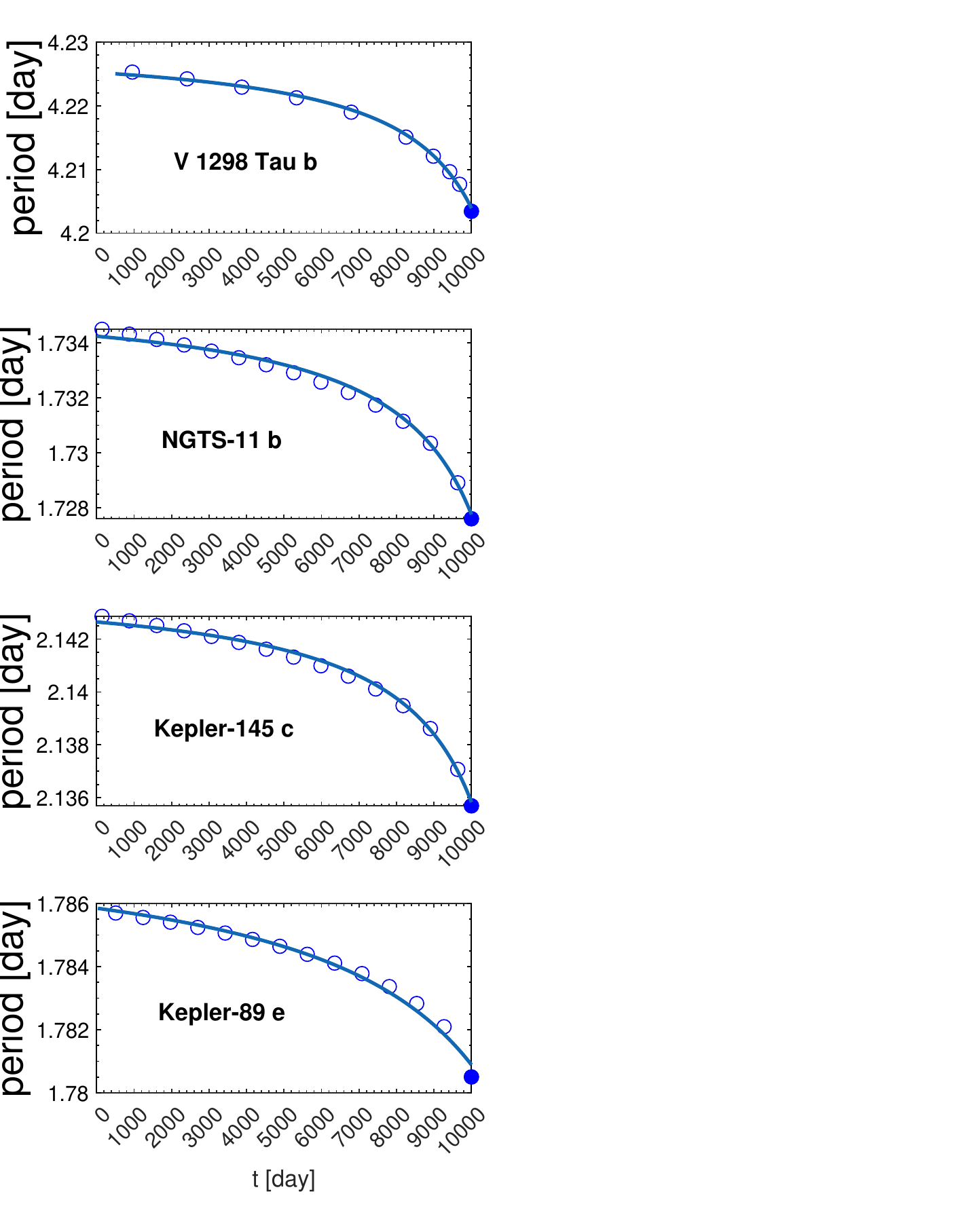}
\caption{The orbital period as function of time in the last $10,000 \days$ before CEE onset for four systems as indicated. Open circles are the results from the \textsc{mesa}-binary simulations while the solid lines are the fittings according to equation (\ref{eq:Orbit}) with the parameters in the last three columns of Table \ref{tab:Table1}. }
\label{fig:P(t)}
\end{figure}

To facilitate comparison with future observations of faint ILOTs we present in Fig. \ref{fig:P(t)} the evolution of the orbital period in the last $10,000 \days$ before CEE onset. 

\cite{Tylendaetal2011} fit a formula for the decrease of the orbital period with time in the last thousands of days before merger of V1309~Sco. 
We fit the same form of the formula but with different parameters. The formula for the orbital period reads  
\begin{equation}
P=A \exp \left( \frac{B}{t-t_0} \right). 
\label{eq:Orbit}
\end{equation} 
We list the value of the parameters in the last three column of Table \ref{tab:Table1}. We take $t=0$ in this fitting to be at $10,000 \days$ before the onset of the CEE (merger), i.e., merger takes place at $t=10,000 \days$ in this formula. 
We note that the value of $t_0$ is larger than the time of merger in this fitting ($10,000 \days$). This is as the fitting of \cite{Tylendaetal2011} to V1309~Sco where $t_0$ is about 700 days after the start of the main peak in the light curve.
The parameters \cite{Tylendaetal2011} fit to equation (\ref{eq:Orbit}) for V1309~Sco are $(A,B,t_0)=(1.4456, 15.29, 10,700)\days$.  
We find that the final orbital shrinkage towards CEE of the cases we study here is qualitatively similar to the observations of V1309~Sco.  

\begin{figure}
\includegraphics[trim= 3.0cm 7.5cm 3.5cm 8.0cm,clip=true,width=0.5\textwidth]{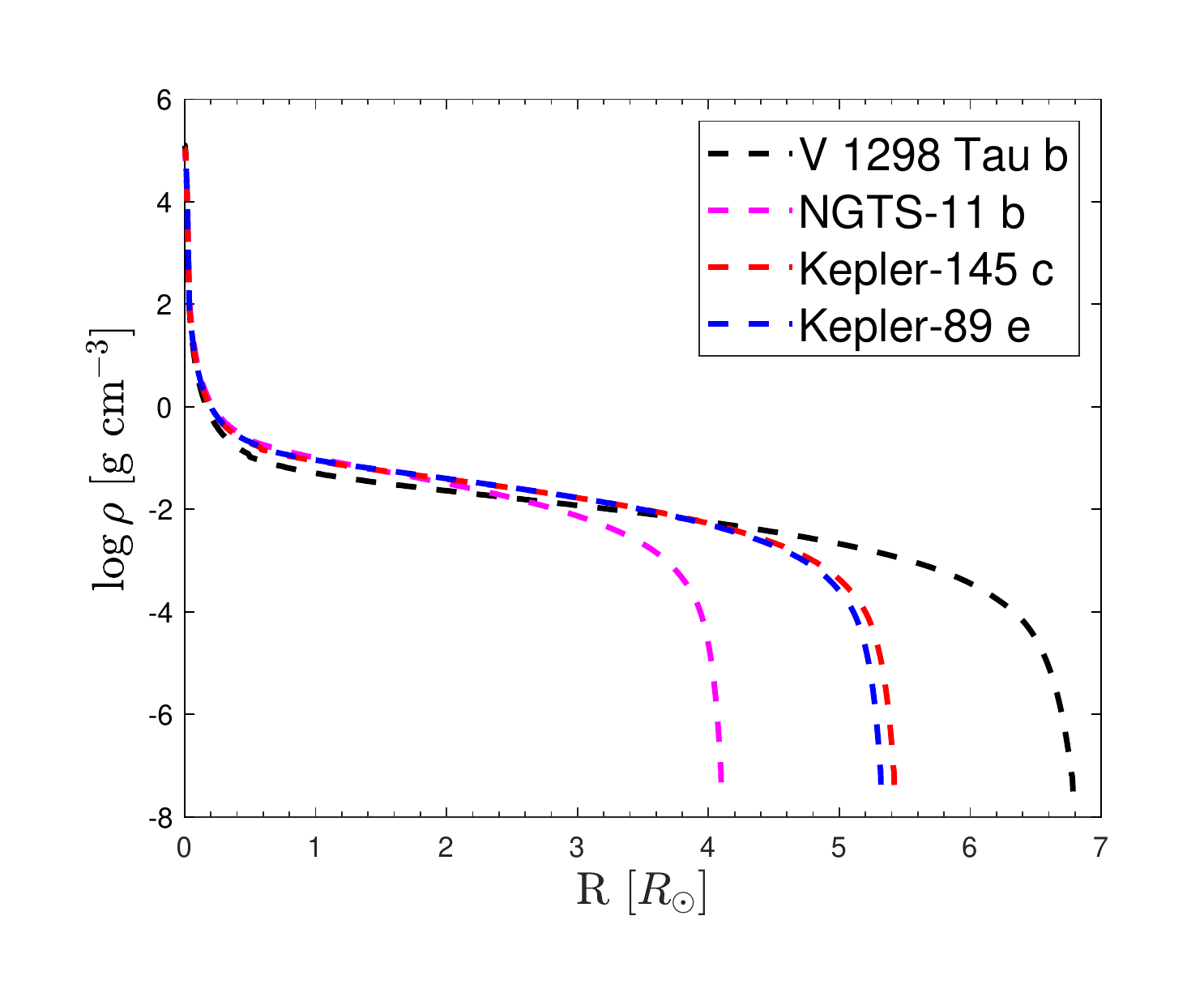}  \\ 
\includegraphics[trim= 3.0cm 7.5cm 3.5cm 8.0cm,clip=true,width=0.5\textwidth]{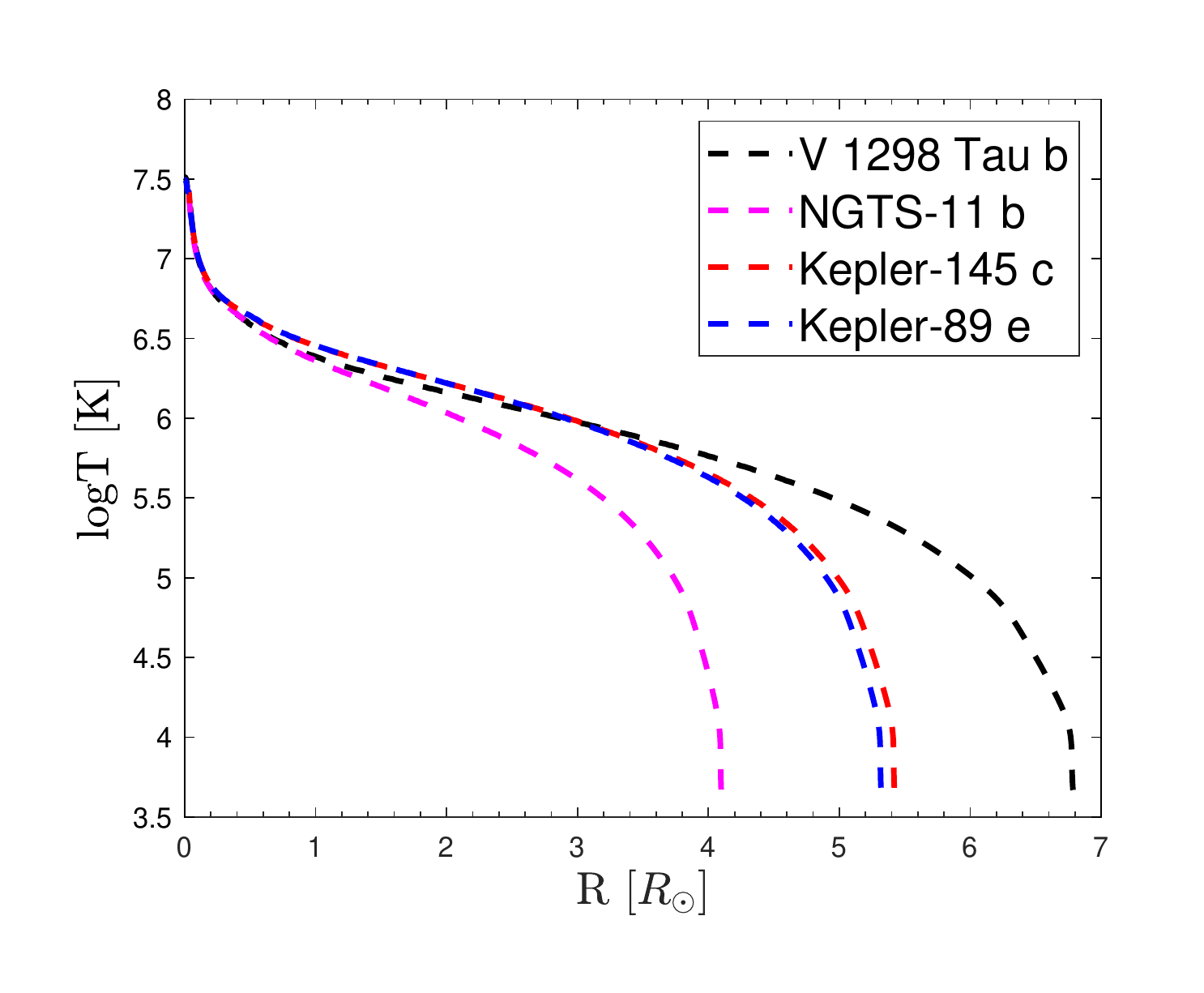}
\caption{Upper panel: Stellar density (in $\g \cm^{-3}$) versus radius at the onset of the CEE for four systems as we list in the inset. Lower panel: The temperature (in K) profiles at the same time. }
\label{fig:logRho_profile}
\end{figure}

The exoplanets we study here have too low masses to eject much of the envelope of their parent stars. Therefore, after the planet enters a CEE, the star is likely to destroy the planet before the planet can reach the core. In principle, the planet destruction can be by tidal interaction with the center of the star or by evaporation. In Fig. \ref{fig:logRho_profile} we present the temperature and density profiles of the stars at the moment they engulf the planet. As the planet average density is $\simeq 1 \g \cm^{-3}$, it will not be tidally destroyed before it is deep inside the star, $r \ll 1 R_\odot$, where the temperature is $\ga 3 \times 10^6 \K$. This is more than an order of magnitude above the virial temperature of the planet. The condition for evaporation, which is for the stellar envelope sound speed to be larger than the escape velocity from the planet, is fulfilled at larger radii than the radii where the tidal destruction condition holds (e.g.,  \citealt{NelemansTauris1998, Soker1998, Krameretal2020}). We therefore expect the planet to be evaporated inside the stellar envelope.
The planet will not influence much the later evolution of the star, e.g., the star will continue very close to its original evolutionary track. 

The remnant of the merger in the case of V1309~Sco is a “blue straggler” star (e.g., \citealt{Ferreiraetal2019}). The planets we study here will have a much smaller effect years after merger, and the star will return to about its pre-merger state. 

\section{ILOT events}
\label{sec:ILOTs}

We here crudely estimate the expected ILOT luminosity and time scale. For that we compare the systems we study to V1309~Sco.
The basic process in our systems is of a planet that enters a CEE and by that ejects mass from the stellar envelope and energise an ILOT. In V1309~Mon it was a low-mass main sequence star that entered the envelope of a more massive star.
From Fig. \ref{fig:P(t)} we see that the systems we study have final orbital periods of $P_{\rm f} \simeq 1.7-4.2 \days$ compared with $P=1.42 \days$ in the case of V1309~Sco \citep{Tylendaetal2011}. \cite{Tylendaetal2011} further estimate that the V1309~Sco binary luminosity was $3-8.6 L_\odot$ and that orbital separation at the merger was of $5.4 - 7.7 R_\odot$. 
Overall, the orbital and stellar properties of the systems we study here are similar to those of V1309~Sco.
The main difference between the systems we study and V1309~Sco is that in the systems  we study the companions are planets rather than a main sequence star. 
\cite{Nandezetal2014} take the stellar masses of the binary progenitor system of V1309~Sco to be $1.52 M_\odot$ and $0.16 M_\odot$, but these are highly uncertain.  
From their simulations \cite{Nandezetal2014} estimate the ejecta mass as $M_{\rm ej} \simeq 0.04-0.05 M_\odot$ (observations give only a lower limit of about $0.001 M_\odot$, e.g., \citealt{Kaminskietal2018}), its energy as $E_{\rm ej} \simeq 1.5 \times 10^{46} \erg$, and that the ejecta takes up to one third of the initial orbital energy. They also find from their simulations that the typical ejecta velocity at large distances is about half the escape velocity from the merging system. These velocities are $\simeq 200 \km \s^{-1}$ as observed \citep{Kaminskietal2015}.   

The total radiated energy, accounting for bolometric luminosity and not only visible (e.g.,\citealt{TylendaKaminski2016}), is $\approx 10^{45} \erg$. Namely, only $\approx 5-10\%$ of the outburst energy of V1309~Sco was carried in radiation.  
In the case of planets this fraction might in some cases be larger because of much lower ejecta mass, which it turn implies that the ejecta is more transparent. 

Consider then a simple scenario by which a planet of mass $M_{\rm p}$ ejects a mass of 
\begin{equation}
M_{\rm ej} \approx 0.05 \left( \frac{M_{\rm p}}{0.16 M_\odot}  \right) M_\odot =
3 \times 10^{-4} \left( \frac{M_{\rm p}}{M_{\rm J}} \right) M_\odot,
\label{eq:Mej}
\end{equation} 
where we scale according to the estimate of \cite{Nandezetal2014} for the ejected mass in V1309~Sco. The ejecta velocity is about the same (or might be somewhat lower), namely, about a half of the escape velocity. Although the energy of the ejecta is much smaller for planets, because of this lower mass the ejecta is less opaque and radiation diffuses out through the ejecta early on. 

Below we assume a spherical mass ejecta. This is a strong assumption since the onset of the CEE ejects mass in a highly-non-spherical geometry (e.g., \citealt{Nandezetal2014}). However, when we use this crude estimate together with the scaling to the observations and simulations of V1309~Sco, we suggest that the results do give an order of magnitude correct estimate of the luminosity and timescale of the event.  

Consider an ejecta expanding with a velocity of $v_{\rm s}$ in a shell of width $\Delta r_{\rm s}$ at radius $r_{\rm s}$ and with a mass of $M_{\rm ej}$. The ejecta starts hot and cools via radiation and adiabatic cooling. The fraction of thermal energy that the radiation carries out of the initial thermal energy is $f \approx t^{-1}_{\rm diff}/(t^{-1}_{\rm diff} + t^{-1}_{\rm exp})$, where $t_{\rm exp} = r_{\rm s} / v_{\rm s}$ is the expansion time of the shell and where the radiation diffusion time from the shell is 
\begin{eqnarray}
\begin{aligned} 
t_{\rm diff} & \simeq \frac{3 \tau \Delta r_{\rm s}}{c} 
\simeq  55   
\left( \frac{M_{\rm ej}}{0.05 M_\odot} \right)
\left( \frac{\kappa}{1 \cm^2 \g^{-1}} \right) 
\\ & \times 
\left( \frac{r_{\rm s}}{5 \times 10^{13} \cm} \right)^{-1}
\left( \frac{\Delta r_{\rm s}}{0.3 r_{\rm s}} \right) \days,
\end{aligned}
\label{eq:tdiff}
\end{eqnarray}
and where $c$ is the light speed, $\kappa$ is the opacity, and $\tau=\rho_{\rm s} \kappa \Delta r_{\rm s}$ is the shell optical depth. 
During this time the gas reach a distance of $r_{\rm s}(55 \days)=  9.5 \times 10^{13} (v_{\rm s}/200 \km \s^{-1})^{-1} \cm$. Namely, the average radius during this time was $\simeq 5 \times 10^{13} \cm$, as the scaling in equation (\ref{eq:tdiff}). 
This might crudely account for the decline time after the peak luminosity of V1309~Sco. 

The actual situation is much more complicated as collision of different ejecta parts can transfer kinetic energy to more thermal energy hence more radiation (e.g., \citealt{Pejchaetal2016a, SokerKashi2016}) and a stellar companion might accrete mass and launch jets to power ILOTs (e.g.,  \citealt{SokerKashi2016, Soker2020ILOTjets}). However, for the purpose of the present crude estimate we proceed with our assumptions. 

In the case of of a planet, the mass is two to three orders of magnitude lower (equation \ref{eq:Mej}), and the equality of photon diffusion time and expansion time is at a radius about ten times smaller, namely, $r_{\rm s} \approx 5 \times 10^{12} \cm$. 
The expanding shell suffers less of adiabatic cooling and a larger fraction of the thermal energy is radiated away. As well, the time scale is ten times shorter for the same expansion velocity. It might be that for planet companions the ejecta velocity and thermal energy are somewhat lower than the simple scaling. However, in that case the shell will suffer less adiabatic cooling as well. 

We take then the ILOT timescale to be the diffusion time when they are both about equal to the expansion time
$\tau_{\rm ILOT} = t_{\rm diff} = t_{\rm exp}$.
Using equations (\ref{eq:Mej}) and (\ref{eq:tdiff}) and $t_{\rm exp} = r_{\rm s} / v_{\rm s}$ we find 
\begin{eqnarray}
\begin{aligned} 
& \tau_{\rm ILOT} \approx 3 
\left( \frac{M_{\rm p}}{M_{\rm J}} \right)^{1/2} 
\left( \frac{\kappa}{1 \cm^2 \g^{-1}} \right)^{1/2} 
\\ &  \times
\left( \frac{\Delta r_{\rm s}}{0.3 r_{\rm s}} \right)^{1/2}
\left( \frac{v_{\rm s}}{200 \km \s^{-1}} \right)^{-1/2} 
 \days. 
\end{aligned}
\label{eq:ILOTtau}
\end{eqnarray}
We assume that the total radiated energy is proportional to the companion mass and we scale by the above mentioned quantities for V1309~Sco. Namely, the total radiated energy is $E_{\rm rad} \approx 10^{45} (M_{\rm p}/0.16 M_\odot) \erg =6 \times 10^{42} (M_{\rm p}/M_{\rm J}) \erg $. Dividing by the ILOT timescale in equation (\ref{eq:ILOTtau}) we find the typical luminosity of the ILOT to be 
\begin{equation}
L_{\rm ILOT}  
\approx 6000 \left( \frac{M_{\rm p}}{M_{\rm J}} \right)^{1/2} L_\odot .    
\label{eq:ILOTL}    
\end{equation}
 
The total radiated energy is a small fraction of the orbital energy at merger $E_{\rm orb} \simeq 3 \times 10^{44} ({M_{\rm p}}/{M_{\rm J}}) \erg$. 
These parameters place the ILOT below the luminosity and energy of classical novae. 
If the unusual outburst of the young stellar object ASASSN-15qi was indeed an ILOT, as \cite{KashiSoker2017planet} suggest, then this zone below novae is already populated. 

The stellar binary interaction most likely forms a bipolar nebula from the ejecta, as observed for example in the cases of the ILOTs V4332~Sgr and V1309~Sco (e.g., \citealt{Kaminskietal2018}) and in CK~Vul (Nova~1670; \citealt{Kaminskietal2021CKVul}).  This might be due to jets that the companion launches as it accretes mass (e.g., \citealt{SokerKashi2016, Soker2020ILOTjets}). We do not expect the planets in the systems we study to form such bipolar structures. 

\section{Summary}
\label{sec:Summary}
 
We followed the evolution of four observed exoplanets to the time that their parent stars engulf them (Fig.\ref{fig:periastron_radius_e_age}). This takes place just after the stars have left the main sequence, i.e., on or near the Hertzsprung gap. Our simulations with \textsc{mesa}-binary show that the final spiralling-in towards forming a CEE is similar to what was observed in the ILOT V1309~Sco (Fig. \ref{fig:P(t)}). 
The main difference between the observed ILOT V1309~Sco and the four systems we study here is that the typical planet mass in the systems we study is about two to three orders of magnitude below the mass of the lower-mass companion in the progenitor of  V1309~Sco (section \ref{sec:ILOTs}). Scaling from V1309~Sco to the systems we study here, we crudely estimate the typical duration and luminosity of the ILOTs to be according to equations (\ref{eq:ILOTtau}) and (\ref{eq:ILOTL}) , respectively.
 
On a broader scope, our study adds to the variety of ILOTs that planets can power as they interact with a more massive companion. We listed some earlier studies of planet-powered ILOTs in section \ref{sec:intro} (\citealt{RetterMarom2003, Retteretal2006, Bearetal2011, KashiSoker2017planet, Kashietal2019Galax}).  Many of these ILOTs have peak luminosity and total energy below those of classical nova, and, therefore, literally speaking they are not `intermediate luminosity'. 
Because of their low luminosity and short duration, planet-driven ILOTs on the Hertzsprung-gap are more difficult to identify. However, because they are simpler than ILOTs driven by stellar companions they can teach us on more energetic ILOTs that are driven by stellar binary systems. They are simpler because they are expected to eject less dust, hence be more transparent, and because, unlike a stellar companion, the planet does not release much energy by accreting mass.

\section*{Acknowledgements}
 
We thank Amit Kashi and an anonymous referee for helpful comments. 
This research was supported by a grant from the Israel Science Foundation (769/20).

\section*{Data availability}

The data underlying this article will be shared on reasonable request to the corresponding author. 

\label{lastpage}
\end{document}